\documentclass[11pt]{article}
\usepackage{a4wide}
\usepackage{epsfig,epsf,graphics}
\usepackage{times}
\usepackage{color}
\usepackage{amsfonts,amssymb,latexsym,amsmath} 
\usepackage{bm}
\usepackage{enumerate}
\usepackage{enumitem}
\usepackage{hhline}
\usepackage{longtable}
\usepackage{array} 
\usepackage{epstopdf}

\newcommand{\al}{&\!\!\!}

\renewcommand{\arraystretch}{1.2}

\newcommand{\email}[1]{\footnote{{\em E-mail address:} \texttt{#1}}}

\begin{document}
\title{ Tetraquarks, hadronic molecules, meson-meson scattering and 
disconnected contributions in lattice QCD }

\author{Feng-Kun Guo$^{\, a,}$\email{fkguo@hiskp.uni-bonn.de} ,
Liuming Liu$^{\, a,}$\email{liuming@hiskp.uni-bonn.de} ,
Ulf-G.~Mei{\ss}ner$^{\, a,b,}$\email{meissner@hiskp.uni-bonn.de} ,
Ping Wang$^{\, c,}$\email{pwang4@ihep.ac.cn} \\ 
   {\it\small$^a$Helmholtz-Institut f\"ur Strahlen- und Kernphysik and Bethe
   Center for Theoretical Physics, }\\
   {\it\small Universit\"at Bonn,  D-53115 Bonn, Germany }\\
   {\it\small$^b$Institut f\"{u}r Kernphysik, Institute for Advanced 
Simulation, and J\"ulich Center for Hadron Physics,}\\
   {\it \small D-52425 J\"{u}lich, Germany} \\
   {\it\small$^c$Institute of High Energy Physics and Theoretical Physics Center 
for Science Facilities, } \\ 
   {\it\small  Chinese Academy of Sciences, Beijing 100049, China}
   }

\maketitle

\begin{abstract}
\noindent
There are generally two types of Wick contractions in lattice QCD calculations 
of a correlation function --- connected and disconnected ones. The disconnected 
contribution is  difficult to calculate and noisy, thus it is often neglected. 
In the context of studying tetraquarks, hadronic molecules and meson-meson 
scattering, we show that whenever there are both connected and singly 
disconnected contractions, the singly disconnected part gives the leading order 
contribution, and thus should never be neglected. As an explicit example, we 
show that information about the scalar mesons $\sigma$, $f_0(980)$, $a_0(980)$ 
and $\kappa$ will be lost 
when neglecting the disconnected contributions.
\end{abstract}

\newpage

It is generally believed that Quantum Chromodynamics (QCD), the fundamental 
theory of the strong interactions, allows for the existence of tetraquark 
states in addition to the normal mesons made of a quark and an antiquark. A 
tetraquark consists of two quarks and two antiquarks. Furthermore, if the 
interaction between two hadrons is strong enough, it should be possible that the 
hadrons can bind together to form a hadronic molecule. Even if the attraction is 
too weak to obtain a bound state, it might still be possible that the 
interaction is resonant. In this case, the resonance generated by the 
interaction is still called a hadronic molecule. Candidates of the tetraquark 
states and/or hadronic molecules among the light mesons include the light scalar 
mesons below 1~GeV, specifically the $f_0(500)$ (or $\sigma$), 
$f_0(980)$, $a_0(980)$ and  $K_0^*(800)$ (or $\kappa$), see  e.g. 
Refs.~\cite{Jaffe:1976ig,Maiani:2004uc,Baru:2003qq,Amsler:2004ps} 
and references therein. In the last decade, due to the continuing world-wide
experimental  efforts, quite a few prominent candidates of tetraquarks and 
hadronic molecules were discovered in the heavy flavor sector. Arguably the most 
intriguing ones are the $X(3872)$~\cite{Choi:2003ue}, whose mass is extremely 
close to the $D^0\bar D^{*0}$ threshold, and the charged heavy quarkonium-like 
states $Z_b(10610,10650)^\pm$~\cite{Belle:2011aa} and the 
$Z_c(3900)^\pm$~\cite{Ablikim:2013mio,Liu:2013dau}.

\begin{figure}[t]
\begin{center}
  \includegraphics[width=0.85\textwidth]{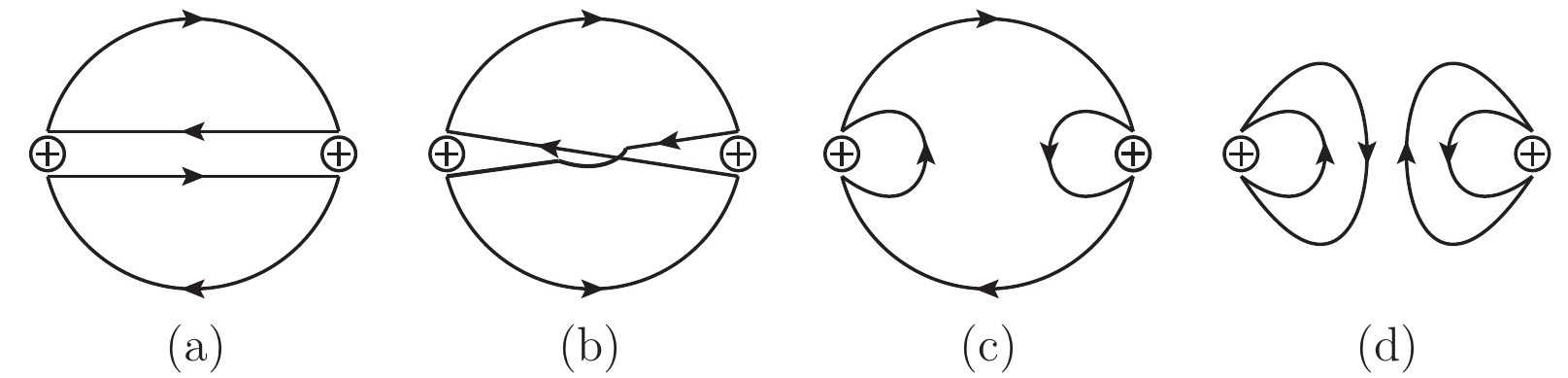}
\caption{Four different types of Wick contractions of the two-point correlation 
function in case of a four-quark interpolating field. The insertion of the interpolating 
field is denoted by $\oplus$. (a) and (b) are connected, (c) is singly 
disconnected, and (d) is doubly disconnected.
}
\label{fig:wickcontractions}
\end{center}
\vspace{-15pt}
\end{figure}

Being intrinsicly nonperturbative, lattice QCD is the ideal approach to study 
hadron spectroscopy including tetraquarks and hadronic molecules. For 
studying tetraquark states in lattice QCD, one calculates the two-point 
correlation function of an interpolating field containing four quarks. The 
basic form of such an interpolating field reads 
\begin{equation}
    \mathcal{O}_{ABCD}^{ij}(x) = \left[\bar q_A(x) \Gamma^i q_B(x) \right] 
\left[\bar q_C (x) \Gamma^j q_D(x) \right],
\label{eq:interfield}
\end{equation}
where the capital indices denote the quark flavor, and $\Gamma^i$ 
are the spin matrices. Notice that each pair of quark and antiquark is 
colorless. This can be done because Fierz transformation allows any 
gauge-invariant four-quark operator to be written in terms of a linear 
combination of products of two colorless quark-antiquark operators. The 
interpolating field used in a lattice calculation can be a linear 
combination of the one given in Eq.~\eqref{eq:interfield} with different spin 
and flavor structures. Depending on the flavor content, in general there are 
four types of Wick contractions in the calculation of the two-point correlation 
function. They are shown in Fig.~\ref{fig:wickcontractions}. 

In lattice calculations, (a) and (b) in Fig.~\ref{fig:wickcontractions} are 
called connected diagrams, and (c) and (d) are the disconnected ones. The 
disconnected contribution involves all-to-all propagators, and is notoriously 
difficult to calculate and often noisy. So far, there have been quite a few 
lattice calculations by different groups on tetraquark states. The simulations 
performed before 2009 were using the quenched approximation, see, e.g., the 
calculations of the light scalar mesons in 
Refs.~\cite{Alford:2000mm,Mathur:2006bs,Suganuma:2007uv,Loan:2008sd, 
Prelovsek:2008rf}. All of them neglect the disconnected diagrams. 
Recently, dynamical calculations in lattice QCD of scalar 
tetraquarks appeared~\cite{Prelovsek:2010kg,Alexandrou:2012rm}, where the 
disconnected diagrams are also neglected despite the fact that unitarity is broken by 
such a procedure. Yet, with the help of various new algorithms and the 
development of computing techniques, e.g. the distillation 
method~\cite{Peardon:2009gh}, there have been lattice calculations including 
disconnected Wick contractions in the context of tetraquarks and meson-meson 
scattering, see, for instance, 
Refs.~\cite{Bali:2011rd,Prelovsek:2013cra,Lang:2012sv,Dudek:2012xn, 
Prelovsek:2013ela}. Energy levels above the meson-meson thresholds which may be 
interpreted as the $\sigma$ and the $\kappa$ were reported in 
Ref.~\cite{Prelovsek:2010kg}. It was shown later in a dynamical lattice 
calculation of $K\pi$ scattering~\cite{Lang:2012sv} that the energy level 
above and near the $K\pi$ threshold is an artifact of neglecting the 
disconnected contributions. However, using similar lattice set-ups as those of 
Ref.~\cite{Alexandrou:2012rm} and also neglecting the disconnected diagrams, the 
ETM Collaboration investigated the isospin 1 and 1/2 channels with a few 
different four-quark interpolating fields. They did not find any state in 
addition to the scattering states in both channels.  The tension between the 
results of the two different groups, as well as the inconsistent results 
obtained in those quenched calculations, stimulate us to scrutinize the validity 
of neglecting the disconnected diagrams. As will be shown in this Letter, as 
long as there are both connected and disconnected contributions, the 
disconnected part is at least as important as the the connected one, or even 
more important depending on the flavor content of the interpolating field. Our 
conclusion can be applied when the correlator in question is a two-point 
correlation function of a four-quark interpolating field. This include studies 
of tetraquarks, hadronic molecules, and meson-meson scattering observables. 

To start, consider the  large-$N$ counting, with $N$ being the number of colors,  
of the various types of contractions, and then present quantitative results on the 
impact of the disconnected diagrams in the sector of the light scalar mesons using 
unitarized chiral perturbation theory.

When the number of colors is large, many interesting qualitative results can be 
obtained~\cite{'tHooft:1973jz,Witten:1979kh}. We will apply the $N$ counting 
rules here to the Wick contractions. Notice that we do not need to assume 
tetraquarks to survive in the large-$N$ limit, which was done in  
Refs.~\cite{Weinberg:2013cfa,Knecht:2013yqa}. What we need is just a $N$ 
counting scheme which is assumed to work for $N=3$ as well. In this way, tetraquarks, 
hadronic molecules and meson-meson scattering can be discussed in a unified way 
with the interpolating field given in 
Eq.~\eqref{eq:interfield}.~\footnote{Hadronic molecules made of two mesons 
cannot exist when $N$ is infinite even if an infinite chain of $1/N$ suppressed 
scattering amplitudes is summed up. This is because the loop expansion of the 
meson-meson scattering coincides with the $1/N$ expansion. An infinite summation 
may be roughly represented by $1/(N-f)$, where $f$ is an $N$-independent factor. 
Apparently, if $N$ is large enough, this summation cannot develop a pole. For 
the behavior of the scalar mesons at a large but not infinite $N$, we refer to 
the review~\cite{Pelaez:2010er} and references therein.} 

Let us analyze the power of $N$ of the different contractions in 
Fig.~\ref{fig:wickcontractions}. Both the cross connected diagram (b) and the 
singly disconnected diagram (c) have only one quark loop, thus they are of order $N$. 
The direct connected diagram (a) and the doubly disconnected diagram (d) have 
two quark loops. However, at least two gluons need to be exchanged to connect 
the two quark loops, since otherwise there is no interaction between the two 
parts and a pole can never emerge. Because the quark-gluon vertex goes as 
$1/\sqrt{N}$, diagrams (a) and (d) should be of order $N^2/(\sqrt{N})^4 = N^0$. 
Therefore, the direct connected and doubly disconnected diagrams are suppressed 
by $1/N$ in comparison with the cross connected and singly disconnected ones. 

We divide the four-quark interpolating fields into different 
types:~\footnote{Since we focus on the correlators with interpolating 
fields of four quarks, we will not discuss another type of tetraquark which is 
excited inside a $\bar q q$ meson as proposed in Ref.~\cite{Knecht:2013yqa}.} 
\begin{itemize}\itemsep0em
    \item Those with one quark having the same flavor as one antiquark, i.e., 
     $\bar q_A q_B \bar q_B q_C$. The possible Wick 
     contractions are given by diagrams (a) and (c) (if $q_C=q_A$, then diagram 
     (d) also contributes). The leading order contribution comes from the 
     singly disconnected diagram (c).
	\item Those with two same quarks (or antiquarks), 
     i.e., $\bar q_A q_B \bar q_C q_B$. The contractions include 
     (a) and (b), both of which are connected, and diagram (b) dominates. If 
     $q_C=q_B$, the singly disconnected diagram (c) will also contribute. It is 
     of the same order as (b), and it is thus not legitimate to neglect it. 
	\item  Those with four different flavors. The 
     two-point correlator only has a contraction of diagram (a) which is 
     connected.
\end{itemize}
From the above large-$N$ counting, we can conclude that as long as the singly 
disconnected contraction contributes, it is of the leading order. Neglecting  
such a contribution will definitely lead to unphysical results which are 
difficult to be related to the real world.

For the light pseudoscalar meson scattering, which is under intense 
investigation in lattice QCD currently (see e.g. 
Refs.~\cite{Beane:2008dv,Lang:2012sv,Dudek:2012xn,Fu:2013ffa,Prelovsek:2013ela} 
and references therein), the same conclusion can be 
obtained without the large-$N$ arguments.
In the low-energy regime of QCD,  the lowest-lying pseudoscalar mesons $\pi$, 
$K$ and $\eta$ can be treated as the (Pseudo-)Goldstone bosons of the spontaneous chiral 
symmetry breaking, and their interactions among themselves and with other 
particles can be described by chiral perturbation 
theory (CHPT)~\cite{Weinberg:1978kz,Gasser:1983yg}. The leading order 
Lagrangian of CHPT only has a single flavor trace. Noticing that the number of 
circles in the Wick contractions is in line with the number of flavor traces, 
only diagrams (b) and (c) can appear at  leading order. Therefore, 
we again arrive at the conclusion that the singly disconnected diagram is 
of leading order, whenever it contributes, and cannot be neglected. As an 
explicit example, we will study the light scalar mesons below 1~GeV using the 
chiral unitary approach.

In the chiral unitary approach as developed in Ref.~\cite{Oller:1997ti}, the 
meson-meson scattering amplitudes derived from the leading order CHPT are put 
in the kernel of the Bethe-Salpeter equation, which is reduced from an integral 
equation to an algebric equation due to an on-shell approximation. The equation 
reads as
\begin{equation}
    T(s) = V(s) [1 - G(s) V(s)]^{-1},
    \label{eq:bse}
\end{equation}
where $V(s)$ is the scattering amplitudes projected to the $S$-wave, and $G(s)$ 
is the normal scalar two-point loop function. For coupled channels, $V(s)$, 
$G(s)$ and $T(s)$ are matrix-valued. With one parameter, which is a 
three-momentum sharp cut-off $q_\text{max}\simeq0.9$~GeV used to regularize the 
divergent scalar loop integral, the scattering data in the isospin $I=0$ and 1 
channels are well described\footnote{Note, however, that better regularization
  methods are by now available. Here, since we are only interested in
  quantitative estimates, we continue using a sharp cut-off.}. 
More interestingly, the lowest-lying scalar mesons 
$\sigma$ (also called $f_0(500)$) and $f_0(980)$ in the isoscalar channel and 
the $a_0(980)$ in the isovector channel can be generated at the right 
positions. We will show that if the disconnected contributions are discarded 
from the scattering amplitudes, the poles of the scalar mesons will be moved 
far away from their physical values. Although extensions of this approach to 
higher orders exist in the 
literature~\cite{Oller:1997ng,Oller:1998zr,Oller:2000fj, GomezNicola:2001as}, it 
is enough to use the simplest formalism in Ref.~\cite{Oller:1997ti} for our 
purpose.

As a first step, one has to separate the disconnected (or connected) 
contributions from the full scattering amplitudes. At  leading order in 
CHPT, i.e. $\mathcal{O}(p^2)$, with $p$ representing a small momentum or a 
Goldstone boson mass, the separation of the different Wick contractions can be 
done by merely extending the normal SU(3) CHPT to SU(4) with one auxiliary 
quark flavor~\footnote{At higher orders, one needs to use partially 
quenched CHPT~\cite{Bernard:1992mk,Sharpe:2000bc} (for reviews, see 
Refs.~\cite{Sharpe:2006pu,Golterman:2009kw}), and examples may be found in 
Refs.~\cite{Sharpe:2006pu,Tiburzi:2009yd,DellaMorte:2010aq,Juttner:2011ur}.}, 
to be denoted by $j$. The meson fields are collected in a $4\times4$ matrix 
\begin{equation}
  \Phi = \begin{pmatrix}
    \eta_u & \pi^+ & K^+ & \phi_{u\bar j} \\
    \pi^- & \eta_d & K^0 & \phi_{d\bar j} \\
    K^- & \bar K^0 & \eta_s & \phi_{s\bar j}\\
    \phi_{j\bar u} & \phi_{j\bar d} & \phi_{j\bar s} & \eta_j 
  \end{pmatrix},
\end{equation}
where $\eta_{f} = \bar f f$ ($f=u,d,s,j$), $\pi^0= \left(\eta_u - \eta_d 
\right)/\sqrt{2} $ and $\eta = \left( \eta_u + \eta_d - 2 \eta_s 
\right)/\sqrt{6} $. We will explain the procedure by an example. Let us 
calculate the connected contribution to the leading order scattering amplitude 
of the process $\pi^0 K^+ \to \pi^0 K^+$, which may be written as
\begin{equation}
    V_{\pi^0K^+\to\pi^0K^+} = \frac12 \left( V_{\eta_u K^+\to \eta_u K^+} + 
    V_{\eta_d K^+\to \eta_d K^+} -2 V_{\eta_u K^+\to \eta_d K^+} \right)~.
\end{equation}
Because the flavor content of the $K^+$ is $\bar s u$, it is obvious that the 
last term in the above equation only has disconnected Wick contractions due to 
the presence of one $\eta_d$. The second term has both connected and 
disconnected contractions. However, the $d$ and $\bar d$ quarks have to be 
contracted within the two $\eta_d$'s, so that the connected contraction 
corresponds to a double-trace in the flavor space. Such a contribution vanishes 
at  leading order of CHPT as discussed above.
\begin{figure}[tb]
\begin{center}
  \includegraphics[width=0.75\textwidth]{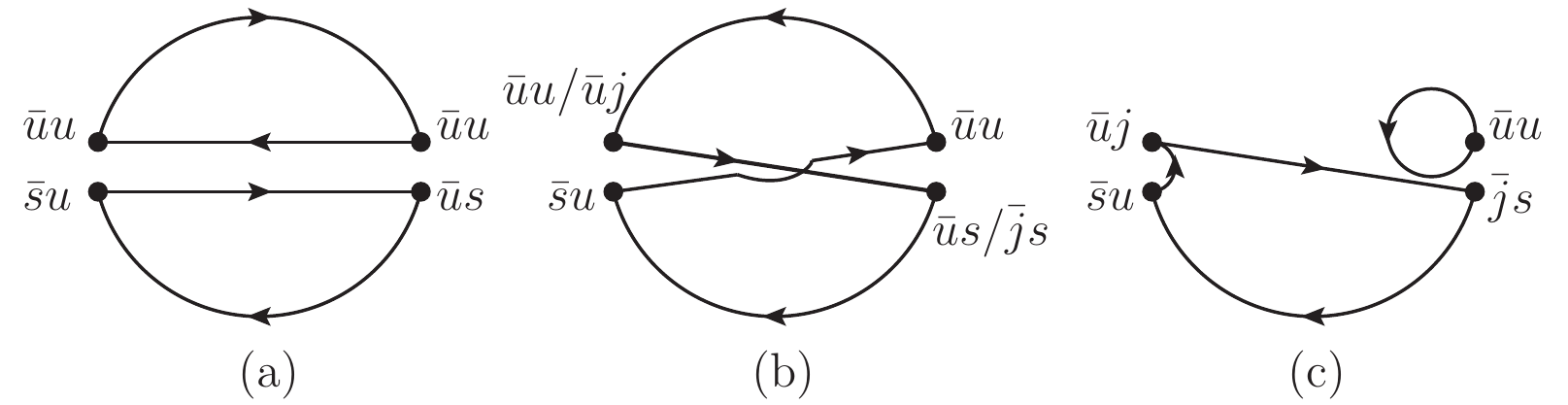}
\caption{The connected Wick contractions for the process 
$\eta_u K^+\to \eta_u K^+$ are given by (a) and (b), while (b) and (c) exhaust 
all possible contractions for the $\phi_{j\bar u}K^+\to\eta_u\phi_{j\bar s}$.
}
\label{fig:etauKp}
\end{center}
\vspace{-15pt}
\end{figure}
There are two different connected contractions for the $\eta_u K^+\to \eta_u 
K^+$, as shown in Fig.~\ref{fig:etauKp}~(a) and (b). Diagram (a) does not 
contribute at the leading order. In addition, this process has disconnected 
contractions including the singly disconnected one. The cross connected diagram 
(b) can be picked out by substituting the auxiliary quark $j$ for one cross line 
of the $u$ quark. To this end, one calculates the scattering amplitude 
of the reaction $\phi_{j\bar u}K^+\to\eta_u\phi_{j\bar s}$. The auxiliary process 
$\phi_{j\bar u}K^+\to\eta_u\phi_{j\bar s}$ has two Wick contractions as shown in 
Fig.~\ref{fig:etauKp}~(b) and (c). However, diagram (c) vanishes at leading 
order due to a double-trace again. The mass of the $j$-quark is required to be 
the same as that of the replaced $u$-quark, and this guarantees that the 
amplitude corresponds to the desired Wick contraction.  Hence, we obtain
\begin{eqnarray}
     V_{\pi^0K^+\to\pi^0K^+}^{(2),\, \text{C}}(s,t,u) \al=\al \frac12    
 \left. V_{\phi_{j\bar u}K^+\to\eta_u\phi_{j\bar s}}^{(2)}(s,t,u) 
 \right|_{m_j=m_u} \nonumber \\
 \al=\al \frac1{4 F^2} \left( s - M_\pi^2 - M_K^2 \right),
 \label{eq:Vpi0Kp}
\end{eqnarray}
where $F$ is the pion decay constant, $M_{\pi(K)}$ are the 
masses of the $\pi$ ($K$), and $s,t,u$ are the conventional Mandelstam variables. 
Here, the superscript ``(2)'' denotes the order $\mathcal{O}(p^2)$, and ``C'' 
represents connected. The connected part in Eq.~\eqref{eq:Vpi0Kp} is clearly 
different from the full leading order amplitude of  $\pi^0K^+\to\pi^0K^+$, 
which is given by $-t/(4F^2)$, and the difference is not negligible. For a 
comparison, both the connected and full scattering amplitudes at 
$\mathcal{O}(p^2) $ for seven most relevant processes are given in 
Table~\ref{tab:amplitudes}. 
\begin{table*}[tbh]
\caption{\label{tab:amplitudes} A comparison of the connected and full 
meson-meson scattering amplitudes, denoted by $V^{(2),\, \text{C}}(s,t,u)$ and 
$V^{(2)}(s,t,u) $, respectively, at $\mathcal{O}(p^2)$.}
\vspace{-15pt}
\begin{center}
\renewcommand{\arraystretch}{1.4}
\begin{tabular}{l c c c }
\hline\hline
Isospin & Processes & $V^{(2),\, \text{C}}(s,t,u)$ & 
$V^{(2)}(s,t,u)$ \\ \hline
$I=0$ & $\pi\pi\to\pi\pi$ & $-\frac1{2F^2} \left( s - 2 M_\pi^2 \right)$ 
      & $-\frac1{F^2} \left( 2s - M_\pi^2 \right)$ \\
      & $K\bar K\to K\bar K$ & 0 & $-\frac3{2F^2}\left(s+t-2M_K^2 \right)$ \\
      & $K\bar K\to \pi\pi$ & 0 & $\frac{\sqrt{6}s}{4F^2}$ \\ \hline
$I=1$ & $\pi\eta\to\pi\eta$ & $\frac1{18F^2}\left(3s-4M_\pi^2-2M_\eta^2\right)$ 
      & $-\frac{M_\pi^2}{3F^2} $ \\
      & $K\bar K\to K\bar K$ & 0 & $-\frac1{2F^2}\left(s+t-2M_K^2 \right)$ \\
      & $K\bar K\to\pi\eta$ & $-\frac{\sqrt{6}}{18F^2} 
\left(3s-M_\pi^2-M_\eta^2         -4M_K^2 \right) $ 
      & $-\frac{\sqrt{6}}{36F^2} \left(9s-M_\pi^2-3M_\eta^2 
        -8M_K^2 \right) $ \\ \hline
$I=\frac12$ & $\pi K \to \pi K$ & $-\frac1{4F^2}\left(s-M_\pi^2-M_K^2 \right)$ 
      & $-\frac1{4F^2}\left(4s+3t-4M_\pi^2-4M_K^2 \right)$ \\ \hline\hline
\end{tabular}
\end{center}
\vspace{-15pt}
\end{table*}

Replacing $V(s)$ in Eq.~\eqref{eq:bse} by the scattering 
amplitudes in Table~\ref{tab:amplitudes} after the $S$-wave projection (an 
additional factor of 1/2 should be multiplied to the $\pi\pi\to\pi\pi$ 
amplitude to account for the identical particles), one can search for poles of 
the $T$-matrix. When the full amplitudes are used, with the three-momentum 
cut-off $q_\text{max}=0.9$~GeV, the pole of the $\sigma$ is 
located at $(0.47-i0.20)$~GeV in the second Riemann 
sheet~\cite{Oller:1997ti}. If the disconnected contributions are neglected, the 
pole will moves to $(0.84-i0.62)$~GeV. It is moved deep in the complex plane, 
and has nothing to do with the physical $\sigma$ pole. Taking into account only 
the connected contractions, the $K\bar K$ channel decouples in the isoscalar 
case, and the $f_0(980)$ pole can not be generated any more. Similar to the case 
of the $\sigma$, when the disconnected contributions are dropped, both the 
$a_0(980)$ and $\kappa$ poles will be moved to far away from their physical 
values, $(1.07-i0.34)$~GeV (in the third Riemann sheet) and $(0.93-i0.54)$~GeV 
(in the second Riemann sheet), respectively. These values should be compared to 
$(1.01-i0.06)$~GeV~\cite{Oller:1997ti} and 
$(0.73-i0.25)$~GeV~\cite{Guo:2005wp} (both in the second Riemann sheet) which 
are obtained when the full $\mathcal{O}(p^2)$ amplitudes are considered
(note that $q_\text{max}=0.51$~GeV given in Ref.~\cite{Guo:2005wp} is used 
for  $\pi K$ scattering).

In summary, we have discussed the contribution of the disconnected Wick contractions 
to the two-point correlation functions of the four-quark interpolating fields. 
The discussion applies to tetraquarks, hadronic molecules and meson-meson 
scattering observables. Using the large-$N$ counting of QCD, we have shown that 
the singly disconnected Wick contraction is always of the leading order whenever 
it contributes. This conclusion is further strengthened for the case of the 
light scalar mesons and the scattering processes involving pseudoscalar mesons 
--- the singly disconnected contraction appears at the leading order of CHPT. 
The connected contribution to the leading order meson-meson scattering 
amplitudes is worked out by introducing an auxiliary quark flavor. Using 
unitarized CHPT, we have shown explicitly that the poles of the light scalar 
mesons move to far away from their physical positions when neglecting the 
disconnected diagrams. Therefore, in the lattice study of tetraquarks, hadronic 
molecules and meson-meson scattering observables, whenever there are both 
connected and singly disconnected contributions, the disconnected part should 
never be dropped.

\medskip

\section*{Acknowledgments}
We are grateful to Marc Knecht, Santi Peris and Andreas Wirzba for valuable 
discussions. This work is supported in part by the DFG and the NSFC through 
funds provided to the Sino-German CRC 110 ``Symmetries and the Emergence of 
Structure in QCD'', the EU I3HP ``Study of Strongly Interacting Matter'' under 
the Seventh Framework Program of the EU, the NSFC (Grant No. 11165005).


\end{document}